\documentclass[aps,physrev,twocolumn,superscriptaddress]{revtex4-2}

\usepackage{amsmath, amssymb}
\usepackage{graphicx}
\usepackage{subfigure}
\usepackage{natbib}
\usepackage[colorlinks=true,linkcolor=blue,urlcolor=black,citecolor=blue]{hyperref}
\begin{document}

\title{Time-reversal assisted quantum metrology with an optimal control}

\author{Da-Wei Luo}
\email{dawei.luo@stevens.edu}
\affiliation{Center for Quantum Science and Engineering and Department of Physics, Stevens Institute of Technology, Hoboken, New Jersey 07030, USA}

\author{Ting Yu}
\email{Ting.Yu@stevens.edu}
\affiliation{Center for Quantum Science and Engineering and Department of Physics, Stevens Institute of Technology, Hoboken, New Jersey 07030, USA}

\date{\today}

\begin{abstract}
We propose a protocol to overcome the shot noise limit and reach the Heisenberg scaling limit for parameter estimation by using quantum optimal control and a time-reversal strategy. Exemplified through the phase estimation,  which can play an important role in  quantum navigation and measurement, we show that the uncertainty arising from a photon number measurement of the system can saturate the assisted Cream\'er-Rao bound, independent of the phase being estimated. In a realistic case with photon loss,  we show that the optimal estimation may still be attainable by optimal control and a projective measurement on an ancilla two-level system  coupled to photonic modes.
\end{abstract}

\maketitle

\section{Introduction}

To achieve increasingly higher precisions of the measurement of physical quantities, the field of quantum metrology has paved the way for next-generation sensing, detection, and high precision measurement technologies~\cite{Thomas-Peter2011a, Giovannetti2004a, Giovannetti2006a, Dorner2009a, Huelga1997a}. Various ingenious approaches have been proposed to beat  the standard quantum limit (SQL)  using experimentally accessible systems, such as trapped ions~\cite{Roos2006a} as `designer atoms', nonclassical atom ensembles~\cite{Pezze2018a} using collective spin systems, and atom chips~\cite{Riedel2010a,Ockeloen2013a} as interferometric probes. Squeezed light has also been demonstrated to overcome the SQL for gravitational wave detections~\cite{ligo2011a,Schnabel2010a}. For example, by using entangled states or squeeze states, it has been demonstrated that the SQL can be surpassed~\cite{Giovannetti2004a}. One important task in quantum metrology is parameter estimation. It is known that, with $N$ unentangled particles to measure a phase, the measurement precision is limited by the SQL at $\Delta \phi \geq 1/\sqrt{N}$. However, the Heisenberg limit $\Delta \phi \geq 1/N$ may be reached when some entangled states are employed. Such limits have been theoretically studied by a common benchmark known as the quantum Fisher information~\cite{Braunstein1996a,Petz2011a,Holevo2011a,Helstrom1969a,Braunstein1994a}, which dictates the lower bounds for the precision attainable via the Cream\'er-Rao bound~\cite{Rao1992a,Cramer1999a}. 

Recently, a quantum metrology protocol based on a unitary transform and its inverse~\cite{Agarwal2022a,Wang2023b,Agarwal2012a} has been demonstrated to  achieve the Heisenberg limit and approach the Cream\'er-Rao bound for certain parameters when  the photon numbers are large. The proposed protocol may offer a quantum advantage for the parameter estimation,  as it provides many different ways to design unitary operations such as a multi-mode squeezing operation (along with its inverse) etc. Motivated by the promising potentials of this reversal strategy, we explore the possibilities of using quantum optimal control algorithms to implement the unitary transform as the forward and reverse time evolutions in a controlled system to saturate the Cream\'er-Rao bound across all parameter regimes, using phase estimation as an example. Noisy environments can make a detrimental impact to the quantum metrology~\cite{Breuer2002a, Giovannetti2011a, Pezze2018b}. For example, the decoherence caused by photon loss or external noises will limit the capacity of quantum reversal operations, and typically the degraded quantum resources such as entanglement and superposition would not be able to achieve the desirable quantum advantage. In these situations, one needs to fully consider the environmental noise effects and employ quantum control mechanisms ~\cite{Chaves2013a,Chin2012a,Rafal2014a,Dur2014a,Kessler2014a,Unden2016a,Lee2023a}. 

The purpose of this paper is to show that, by combining unitary transformations, engineered quantum states, and control mechanisms, we can realize the promised performance of quantum metrology in a wide range of scenarios including the cases where the detrimental environmental noises are present. It should be noted our approach for the parameter estimation is applicable to both close and open quantum systems, the latter may be coupled to an external environment, such that the photon loss is inevitable. To achieve our goal, we will consider applying an optimal control strategy to a system of two photonic modes and an ancilla two-level system (TLS). The advantage of including such an ancilla TLS is that, it may allow us to use external control fields to modulate the TLS and its coupling to the photonic modes to create an entangled state as a resource state for the quantum metrology.  Such generated resource states may render our protocol optimally operational reaching the Heisenberg scaling while saturating the Cream\'er-Rao bound. The TLS, as an ancilla, is useful for detecting photon loss. By combining a projective measurement on the TLS with additional constraints on the optimal control, we will show how the TLS can be used to compensate the photon loss and recover the ideal performance in an open system setting. Our results has established the quantum meterological advantage for a wide range of physical settings, leveraging our optimal control operations as a useful information restoration method.

\section{Quantum metrology with optimal control and time reversal} 

\begin{figure}[b]
    \centering
    \includegraphics[width=.4\textwidth]{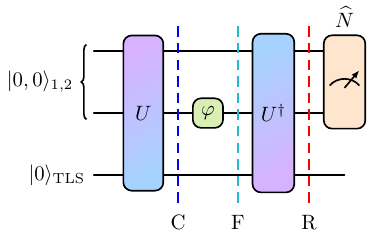}
    \caption{(Color online) Schematic of the phase estimation protocol. Starting with an initial state where the two photon modes are in the vacuum state and the two-level system (TLS) in the ground state $|0 \rangle$, we design a unitary evolution $U$ to estimate the phase parameter $\varphi$ applied to the mode $2$ by a photon count after a time-reversal evolution.}\label{fig_sch}
\end{figure}

The protocol of phase estimation is displayed in Fig.~\ref{fig_sch}, motivated by the squeeze-then-reverse protocol proposed in ~\cite{Agarwal2022a,Agarwal2012a,Wang2023b}. Starting with a two-mode vacuum state and a two-level system in the ground $|0 \rangle$, we evolve the system according to a unitary propagator $U$ of an actively controlled quantum system (stage C in Fig.~\ref{fig_sch}, as an initialization of the setup). Then, the phase parameter $\varphi$ to be measured is applied to the second photonic mode, $a_2 \rightarrow a_2 \exp(i \varphi)$, $a_2$ being the annihilation operator for the second mode. The composite system undergoes a time reversal process according to $U^\dagger$, yielding the final state of the protocol (stage R in Fig.~\ref{fig_sch}). We then measure the photon number of the two modes as the output. The main goal is to design a system Hamiltonian such that the unitary propagator $U$ yields a state that can achieve the desired Heisenberg scaling for the parameter estimation as the number of photons $N$ of the resource state (at stage C) grows.

Note that the use of active controls has been studied in various contexts. For example, using a controlled sequential scheme~\cite{Hou2019a}, a precision near the Heisenberg limit has been experimentally verified. Other control schemes such as feedback controls and machine optimization have been shown to be useful in enhancing quantum fisher information~\cite{Xu2021a,Fallani2022a,Lin2021a,Haine2020a}. In addition, variational-based strategies tuning the quantum controls and initial states to combat quantum noises have also been reported~\cite{Yang2022a,Zhai2023a}.  It should be noted that finding a desirable control scheme for the phase estimation is a daunting task since the parameter to be estimated is an unknown quantity.  We will show how,  by an optimal control scheme to design a parameter-independent unitary propagator (and its reverse operation) to reach the Heisenberg limit for the entire range of the unknown parameter.

To evaluate the effectiveness of our protocol, we use the common benchmark known as the quantum Fisher information~\cite{Braunstein1996a,Petz2011a,Holevo2011a,Helstrom1969a,Braunstein1994a}, which dictates the lower bounds for the precision attainable via the Cream\'er-Rao bound~\cite{Rao1992a,Cramer1999a}.
Hence, we will then program the optimal control scheme such that the Cream\'er-Rao bound can be saturated across the parameter regime $\varphi \in [-\pi, \pi)$.

To start with, we will consider the photonic entangled NOON state $(|N0 \rangle + |0N \rangle )/\sqrt{2}$, which has been known to show exceptional quantum enhancements, and achieve `super resolution'~\cite{Israel2014a,Rehman2022a,Hosler2013a,Hong2021a,Dowling2008a,Giovannetti2004a,Okamoto2008a,Resch2007a,Xiang2010a}. It should be noted that, however,  the NOON state may be susceptible to photon losses~\cite{Joo2011a,Huver2008a,Nielsen2023a,Dorner2009a}. In what follows, we will show that our optimal control based protocol can also deal with the photon loss cases.
To construct the optimal control, we need to first determine the set of initial and target states, where the goal of the optimal control is to find a optimal field(s) $c_i(t)$, such that the time-dependent Hamiltonian
\[
	H(t) = H_0 + \sum_i c_i(t) H_i
\]
can drive the initial states to their respective target states in a prescribed time $t$, where $H_0$ is the time-independent uncontrolled part of the Hamiltonian, $H_i$ is the time-independent Hamiltonian describing the control strategy, and $c_i(t)$ is the corresponding time-dependent control functions.

Here, we use the entangled NOON state $\left(|N0 \rangle + |0N \rangle\right)/\sqrt{2}$ as one of the target states of the quantum optimal control. It has been shown that using an ancilla two-level system (TLS), one can generate arbitrary photonic states by modulating the TLS and controlling the interaction between them~\cite{Law1996a}. It has been later shown that with multiple fields, similar goals can be achieved using three-level atoms~\cite{Drobny1998a}, whereas for some states TLS would also suffice~\cite{Strauch2012a,Sharma2016a}. To implement the unitary gate for the optimal control, we consider a system of two-mode cavity coupled to a TLS, with Hamiltonian

\begin{align}
    & H(t) = \sum_{j=1,2} \omega_j a_j ^\dagger a_j + H_{TLS}(t) + H_I(t) \nonumber \\
    & H_{TLS}(t) = f_x(t) \sigma_x + f_z(t) \sigma_z \nonumber \\
    & H_I(t) = \sum_{j=1,2} g_j(t) \left( \sigma^+ a_j + \sigma^- a_j ^\dagger \right)
\end{align}
where $H_{TLS}$ is the Hamiltonian for the controlled TLS, $H_I$ denotes the interaction between the TLS and the two photonic modes, $\sigma_{x,z}$ are the Pauli operators, $f_{x(z)}$ are the fields applied to the TLS along the $x(z)$ axis, $a_j$ is the annihilation operator for the $j$-th mode, $g_j(t)$ signifies the coupling strengths between the TLS and the photonic modes, and $\sigma^{+(-)}$ are the spin raising(lowering) operator for the TLS. We will use the Hilbert space $TLS \otimes F_1 \otimes F_2$ where $F_{1(2)}$ denotes photonic mode $1(2)$ throughout this paper.

\begin{figure}
    \centering
    \includegraphics[width=.45\textwidth]{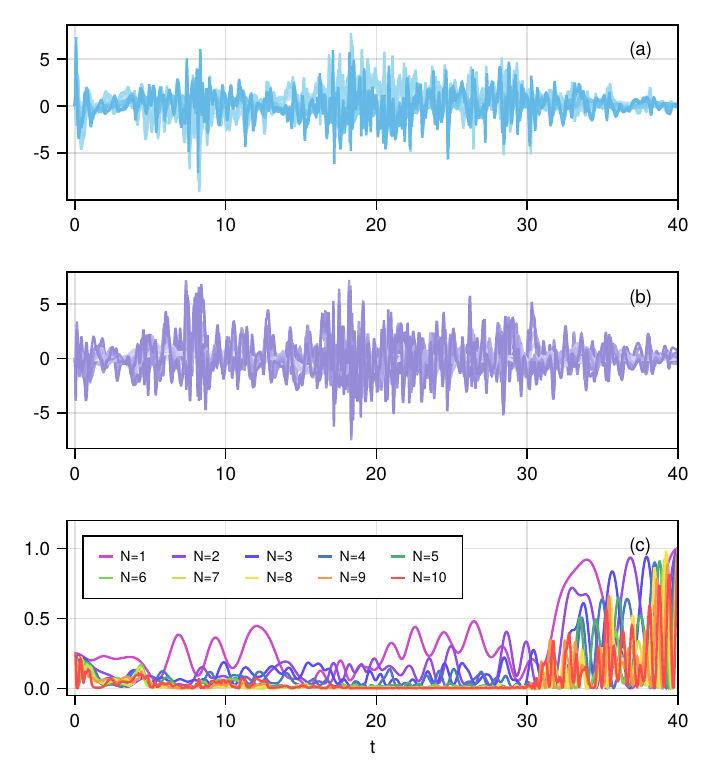}
    \caption{(Color online) Control fields and the corresponding fidelity dynamics, for $N=1\ldots 10$ of the target NOON states. Panel (a) shows the interaction strengths $g_{1,2}(t)$ and Panel (b) shows the frequency modulation $f_{x,z}(t)$. Panel (c) shows the fidelity dynamics. The control runtime is chosen to be $T_f=40$.}\label{fig_ctrlf}
\end{figure}

A reliable generation of an entangled NOON state is very useful for experimental implementations. Here, using an ancilla TLS, we have demonstrated an alternative way to generate the resource state for the quantum metrology task, through controlling the frequencies of the TLS and the couplings between the TLS and the two mode fields. To find the suitable control fields,  we use a gradient based optimal search algorithm known as the Krotov method~\cite{Tannor1992h,Reich2012v,Konnov1999b}. As an iterative algorithm, it has the advantage of being monotonically convergent and does not require a line search. The Krotov algorithm may be used to find the time-dependent fields that drive one initial state to one target state as state preparation or transfer, but it may also be used to find the control fields that evolves multiple initial states to their respective target states, a common goal in quantum gates engineering. Here, we will use the latter to find the control fields to realize the unitary propagator $U = \mathcal{T}\exp[-i\int ds H(s)]$. Mathematically, the Krotov's method is derived to minimize a functional $J$,

\begin{align}
    J\left[s, \{c^{(i)}_l(t)\}\right] = J_T(s) + \sum_l \int_0^T g(\{c^{(i)}_l(t)\}) \label{eq_kctrl_j},
\end{align}
where $s=\{| \varphi^{(i)}_j(t)\}$ is the set of wave functions at time $t$ for the $j$-th initial state at the $i$-th iteration, and $\{c^{(i)}_l(t)\}$ is the set of control functions at iteration $i$. By a clever mathematical trick, the Krotov algorithm decouples the interdependence between the control field and the quantum state's evolution, allowing one to optimize the functional iteratively with some initial guess control field.
The function $J_T$ is the main part of the functional~\eqref{eq_kctrl_j} and for our purpose, $J_T$ is taken to be the infidelity of the evolved state and the target state~\cite{Tannor1992h,Reich2012v,Konnov1999b}
\begin{align}
	J_T(s) = 1 - \sum_j |\langle \phi^{(f)}_{j} | \varphi^{(i)}_j(T) \rangle|^2/N_{c},
\end{align}
where $|\phi^{(f)}_{j}  \rangle$ is the $j$-th target final state, with $N_{c}$ initial-target state pairs. The function $g$ tracks the running cost of the control fields, and is usually taken in the form of
\begin{align}
    g(\{c^{(i)}_l(t)\}) = \frac{\lambda_{a,l}}{S_l(t)}(\Delta c_l^{(i)}(t))^2,
\end{align}
where $\lambda_{a,l}>0$ is an inverse step-size, $\Delta c_l^{(i)}(t)= c_l^{(i)}(t) - c_l^{(i-1)}(t)$ is the difference of the control function between the current and last iteration, and $S_l(t) \in [0,1]$ is an update shape function. The control pulse can then be updated iteratively using
\begin{align}
    \Delta c_l^{(i)}(t) = \sum_k \frac{S_l(t)}{\lambda_{a,l}} \mathrm{Im} \left[\left\langle \chi^{(i-1)}_k(t)\left|\frac{\partial H^{(i)}}{\partial c^{(i)}_l(t)} \right| \varphi^{(i)}_k(t)\right\rangle\right],
\end{align}
where $| \chi_k^{(i)}(t) \rangle $ evolves `backwards' according to $H^\dagger(t)$, with boundary condition at the final $T_f$ as $|\chi^{(i-1)}_k(T_f) \rangle = - \partial J_T(T_f)/\partial \langle \varphi_k^{(i-1)} (T_f)|$.

With the goal of saturating the Cream\'er-Rao bound in mind, we first observe that the state after the external phase $\varphi$ is applied (state F in Fig.~\ref{fig_sch})

\begin{align}
    |\psi_F \rangle = |0 \rangle \left(|N0 \rangle + e^{iN\varphi}|0N \rangle\right)/\sqrt{2} \label{eq_psif}
\end{align}
may be expanded by two basis NOON states $|0 \rangle \left(|N0 \rangle \pm |0N \rangle\right)/\sqrt{2}$. Therefore, in addition to driving the initial vacuum states to the NOON state
\begin{align}
    |\psi_{i,1} \rangle = |0 \rangle |0 0\rangle &\xrightarrow{U}
    |\psi_{t,1} \rangle = |0 \rangle \left[ |N 0 \rangle + |0 N \rangle \right] / \sqrt{2},
\end{align}
where the subscript $i(t)$ denotes the initial (target) state, we can set up another initial-target state pair for the optimal control
\begin{align}
    |\psi_{i,2} \rangle = |0 \rangle |0 N_x \rangle &\xrightarrow{U}
    |\psi_{t,2} \rangle = |0 \rangle \left[ |N 0 \rangle - |0 N \rangle \right] / \sqrt{2},
\end{align}
where $N_x \neq 0$ and can be chosen as $N_x=N$ without loss of generality. Accordingly, the final state at stage R may now be given as
\begin{align}
    |\psi_R \rangle &= U^\dagger U_\varphi U |0 \rangle |0,0 \rangle \nonumber \\
    &= \frac{1+e^{iN \varphi}}{2} |0 \rangle |0 0\rangle + \frac{1-e^{iN \varphi}}{2} |0 \rangle |0 N_x \rangle,
\end{align}
where $U_\varphi=\exp[i\varphi a_2 ^\dagger a_2]$ denotes the application of the external phase. A common benchmark for quantum metrology is the Fisher information, which dictates the lower bound for the uncertainty via the Cream\'er-Rao bound, $\delta^2 \geq 1/mF$, where $\delta$ is the uncertainty, $m$ the number of experimental realizations and $F$ is the Fisher information, defined as $F=\mathrm{tr}[\rho_\phi L^2]$, for a parameterized state $\rho_\phi$ and $L$ is the symmetric logarithmic derivative operator $\partial \rho_\phi/\partial \phi=\{L,\rho_\phi\}/2$.

It can be derived that the quantum Fisher information associated with the output parameterized state $|\psi_R \rangle$ is~\cite{Liu2019a}
\begin{align}
    F_\varphi = 4 \left[\langle \partial_\varphi \psi_R | \partial_\varphi \psi_R \rangle - \left|\langle \partial_\varphi \psi_R| \psi_R \rangle \right|^2\right] = N^2,
\end{align}
which shows Heisenberg scaling with the photon number $N$ of the resource state independent of the parameter $\varphi$. The uncertainty of the phase estimate based on the photon number counting may also be analytically derived, with
\begin{align}
    \langle \hat N \rangle &= N_x \sin^2 \left[\frac{N \varphi}{2}\right], \quad
    \langle \hat N^2 \rangle &= N_x^2 \sin^2 \left[\frac{N \varphi}{2}\right],\label{eq_n_n2}
\end{align}
which leads to
\begin{align}
    \delta \varphi &\equiv \frac{\Delta N}{\left| d \langle \hat N \rangle/d\varphi \right|} = \frac{N_x}{2} \left| \sin(N \varphi) \right| \bigg/ \left|\frac{ N \cdot N_x }{2} \sin(N \varphi)\right| \nonumber \\
    &= 1/N, \label{eq_uncc}
\end{align}
where $(\Delta N)^2 = \langle \hat N^2 \rangle -\langle \hat N \rangle^2$ is the variance of the photon count. Thus, we demonstrate that with our proposed quantum metrology protocol, the uncertainty remains a constant and one can saturate the Cream\'er-Rao bound \emph{independent} of the parameter $\varphi$, and have a Heisenberg scaling with the photon number $N$ in the resource NOON state.

\begin{figure}
    \centering
    \includegraphics[width=.4\textwidth]{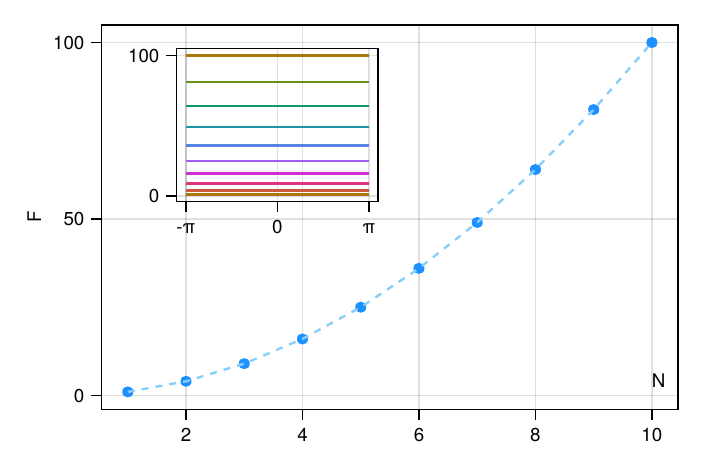}
    \caption{(Color online) Scaling behavior of the quantum Fisher information. The blue dots are from numerical simulation, and the dashed line depicts the theoretical Heisenberg scaling $N^2$. Inset: Fisher information as a function of the parameter $\varphi$ for $N=1 \ldots 10$ (bottom to top).}\label{fig_fisher}
\end{figure}

To numerically study the properties of the proposed protocol, we solve for the control fields for $N=1 \ldots 10$. In Fig.~\ref{fig_ctrlf} we show corresponding control fields and the dynamics of the average control fidelity $|\langle \psi(T_f) | \psi_T \rangle|^2$, where $| \psi(T_f)\rangle$ is the state at the end of the control runtime $T_f=40$, $|\psi_T \rangle$ is the ideal control target, and the average is taken over all initial-target state pairs. The control algorithm is set to a goal of reaching infidelity $1-F \sim 10^{-4}$, and we set $\omega_1=1$, $\omega_2=2$ for the photonic modes.

\begin{figure}
    \centering
    \includegraphics[width=.4\textwidth]{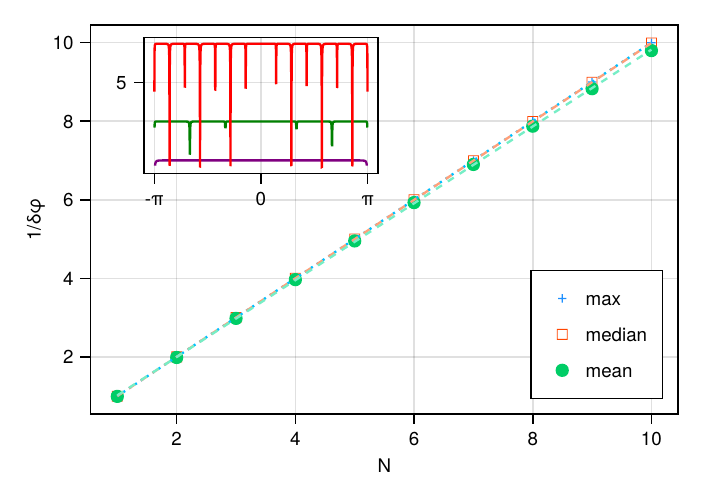}
    \caption{(Color online) Scaling behavior of the inverse-uncertainty $1/\delta \varphi$, for $N=1 \ldots 10$. A linear scaling of the maximum, median and mean inverse uncertainty $1/\delta \varphi$ may be observed, for $\varphi \in [-\pi, \pi]$ taken from the inset with $6000$ evenly spaced points. Inset: the inverse uncertainty $1/\delta \varphi$ as a function of $\varphi$ for $N = (1, 3, 7)$ (bottom to top) as examples (other $N$ values show similar behaviors). Numerical instabilities around $0/0$ are excluded. The drops around $i\pi/N$ for integer $i \in \mathbb{Z}$ are due to both numerical instabilities close to $0 / 0$ (see Eq.~\eqref{eq_uncc}) and the optimal control not achieving the ideal unit fidelity.}\label{fig_duncer}
\end{figure}

The resulting Fisher information and uncertainty may also be numerically obtained. It can be verified that the Fisher information is $\varphi$-independent and shows Heisenberg scaling $F=N^2$ with the number of photons, shown in Fig.~\ref{fig_fisher}. As for the uncertainty $\delta \varphi$, there exists some numerical instabilities around $\sin(N \varphi )$ due to the division near $0/0$ (see Eq.~\eqref{eq_uncc}) and the control being non-ideal fidelity. We show the maximum, median and mean values of the inverse uncertainty in the lower panel of Fig.~\ref{fig_duncer}, where a linear scaling can be observed. Least-square fitting shows that the median inverse fidelity $1/\delta \varphi$ scales near $0.9989 N$, the mean inverse fidelity scales near $0.9782 N$ and the maximum scales close to the ideal $N$, taken with $6,000$ evenly spaced $\varphi \in [-\pi,\pi]$.

\section{Open system considerations: photon loss and its remedy}

In this section, we will consider the case where the resource NOON state may be subject to an environmental noise which can lead to the photon loss. The active control strategy we used before can be adjusted to account for the finite photon losses. However, in this case, an additional measurement on the TLS is needed and the measurement process might need to be repeated depending on the output results. To model the photon loss due to the interaction with an environment, we consider a phenomenological Lindblad Markov master equation~\cite{Breuer2002a}
\begin{align}
    \frac{d\rho}{dt} = \sum_{i=1,2} L_i \rho L_i^\dagger-\frac{1}{2}\{L_i ^\dagger L_i,\rho\},
\end{align}
where $L_i = \lambda_i a_i$ and $\lambda_i$ denotes the coupling strength. This master equation may be unravelled as a series of discrete ``quantum jumps''~\cite{Dum1992a,Plenio1998a,Carollo2003a}. For a pure initial state $|\xi_0 \rangle$, the state at the next time-step $dt$ may be decomposed as~\cite{Carollo2003a}
\begin{align}
    \rho(dt) &\approx \sum_{j=0}^2 K_j |\xi_0 \rangle\langle \xi_0| K_j ^\dagger \nonumber \\
    &= p_0 |\chi_0 \rangle \langle \chi_0| + \sum_{j=1,2} p_j |\chi_j \rangle \langle \chi_j|, \label{eq_rhoj}
\end{align}
with $K_0 = (1-iH_{\rm eff}dt)$ denotes the Kraus operator associated with the `no-jump' state $|\chi_0 \rangle$ with the non-Hermitian effective Hamiltonian $H_{\rm eff} = -i\sum_j L_j ^\dagger L_j/2$, while the `jump' states Kraus operators are given by $K_j = \sqrt{dt} L_j$, occurring with probability $p_j = \langle \xi_0|L_j ^\dagger L_j|\xi_0 \rangle$, and $p_0 = 1 - p_1 - p_2$. Here, we assume the unitary control is ideal, and the photon loss occurs at the phase acquisition stage F in Fig.~\ref{fig_sch}. It may be seen that with photon loss $L_j=\lambda_j a_j$, the NOON state $|\psi_F \rangle$ in Eq.~\eqref{eq_psif} is an eigenstate of the no-jump operator. Without loss of generality, we assume that at most $1$ photon is lost. This will allow us to decompose the decayed state after phase acquisition $\rho_\varphi$ in the form of Eq.~\eqref{eq_rhoj}, with $|\chi_0 \rangle = |\psi_F \rangle$, $|\chi_1 \rangle = |0 \rangle |N-1,0 \rangle$ and $|\chi_2 \rangle = |0 \rangle |0, N-1 \rangle$. When the state is subject to the photon loss, we can also calculate the Fisher information and the uncertainty. For mixed states $\rho$ with support $S$ where $\rho = \sum_{i \in S} \lambda_i |\lambda_i \rangle\langle \lambda_i|$ as eigen-decomposition with $\lambda_i \neq 0$, the Fisher information is given by~\cite{Liu2019a}
\begin{align}
    F_\varphi =& \sum_i \frac{(\partial_\varphi \lambda_i)^2}{\lambda_i}
    + \sum_i 4 \lambda_i \langle \partial_\varphi \lambda_i | \partial_\varphi \lambda_i \rangle \nonumber\\
    & - \sum_{i,j} \frac{8 \lambda_i \lambda_j}{\lambda_i + \lambda_j} |\langle \partial_\varphi \lambda_i | \lambda_j \rangle|^2
\end{align}
where the summations are over the support $S$. The Fisher information for the 1-photon decayed state is scaled by the no-jump probability, $F_\varphi=p_0N^2$. The variance may also be derived, 
\begin{align}
    \delta \varphi = \sqrt{\frac{2-p_0 (1+\cos N \varphi)}{ N^2 p_0(1-\cos N\varphi)}}. \label{eq_loss_unc}
\end{align}
It should be noted that the variance is no-longer $\varphi$-independent. It is easily seen that, at $\cos(N \varphi) = -1$, it takes the minimal value $1/(N\sqrt{p_0})$ so the $1/N$ scaling still holds and saturates the Cream\'er-Rao bound, albeit only for specific $\varphi$'s.

We now show how to adapt the protocol such that the ideal parameter-independent uncertainty as the closed system may be recovered in the case of photon loss. Since with the controlled time reversal, the `no-jump' state $|\chi_0 \rangle$ will be in the subspace $\mathcal{H}_{T0}$ where the TLS is in the $|0 \rangle$ state, we can add two additional constraints to the control goal, such that states with photon loss will be sent to the subspace $\mathcal{H}_{T1}$ where the TLS is in the $|1 \rangle$ state
\begin{align}
    |\psi_{i,3} \rangle = |1 \rangle |N0 \rangle &\xrightarrow{U} |\psi_{T,3} \rangle = |0 \rangle |N-1,0\rangle, \nonumber\\
    |\psi_{i,4} \rangle = |1 \rangle |0N \rangle &\xrightarrow{U} |\psi_{T,4} \rangle = |0 \rangle |0,N-1 \rangle.
\end{align}
In this case, the state after the time reversal is
\begin{align}
    \rho_R &= U^\dagger \rho_\varphi U \nonumber \\
    &= p_0 |\psi_R \rangle\langle \psi_R| + p_1 |\psi_{i,3} \rangle\langle \psi_{i,3}| + p_2 |\psi_{i,4} \rangle\langle \psi_{i,4}|,
\end{align}
where $|\psi_R \rangle \in \mathcal{H}_{T0}$ is the same as the closed system case, corresponding to the no-jump scenario, while states with photon loss evolves to $|\psi_{i,3(4)} \rangle \in \mathcal{H}_{T1}$.

After the time-reversal, we follow up with a positive operator-valued measure (POVM) on the TLS. The measurement proceed only if the TLS is found to be in the $|0 \rangle$ state, in which case $\rho_{R} \rightarrow M_0 \rho_{R} M_0^\dagger / \mathrm{tr}(\cdot) = |\psi_R \rangle\langle \psi_R|$, where $M_0 = |0 \rangle\langle 0|_{\rm TLS}$. In this case, the ideal closed system results can be recovered. Otherwise, we may discard the result and restart the measurement if the TLS is found to be in the $|1 \rangle$ state. States with more than $1$ photon loss may also be sent to the $\mathcal{H}_{T1}$ subspace with the active control in an ad-hoc fashion.

\begin{figure}
    \centering
    \includegraphics[width=.4\textwidth]{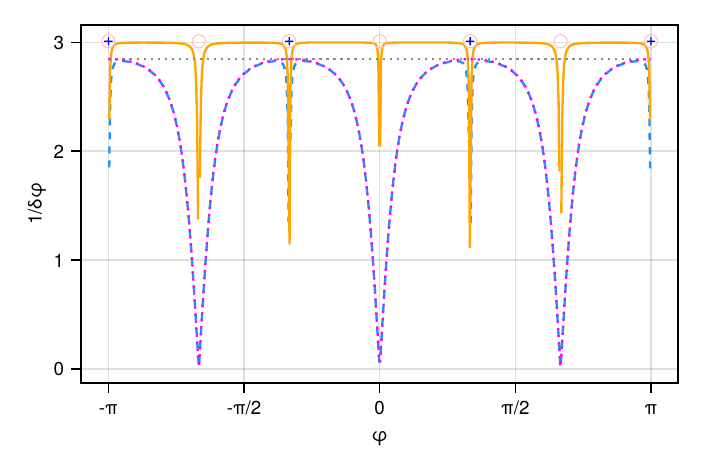}
    \caption{(Color online) Uncertainty under the photon loss as a function of the parameter $\varphi$. The blue dashed line shows the inverse fidelity under the photon loss using numerical simulations, and the pink dotted lines are derived analytically. The orange line shows the recovered uncertainty after a POVM of the TLS. The black dashed line shows the upper bound under the photon loss, $N\sqrt{p_0}$. Points with numerical instabilities near $0/0$ are excluded, marked by pink circles for the POVM case and blue $+$ for the photon loss case.}\label{fig_ploss}
\end{figure}

In Fig.~\ref{fig_ploss} we show numerical results using the adapted control and POVM, choosing $N=3$, $p_0=0.9$, and $p_1=p_2=0.05$. The blue dashed line shows the inverse uncertainty as a function of the parameter $\varphi$ obtained from numerical simulation of the control, in agreement with the pink dotted line showing the analytical values Eq.~\eqref{eq_loss_unc}. The black dotted lines shows the upper bound $N\sqrt{p_0}$ for the decayed state. The orange solid line shows the recovered performance following POVM, $1/\delta \varphi=N$. Therefore, at the cost of an additional POVM on the TLS and needing to restart the measurement with some probability ($1-p_0$), we can recover the closed system performance even with photon losses. When the system-bath interaction is weak or the time scale of the phase acquisition is short, the probability of the no-jump scenario $p_0$ will be close to $1$. In such scenarios, the probability of needing to discard the results and restart the measurement can be lowered.

\section{Discussion and conclusion}
By engineering a two-mode cavity with a two-level system as an ancilla, we are able to design a unitary propagator to generate a photonic entangled state. With a time-reversal operation, we show that the parameter estimation of a phase parameter applied to one of the modes can overcome the shot noise limit and allows for the quantum Fisher information to reach the Heisenberg scaling with the photon number $N$ of the entangled state. Moreover, by a simple photon-number counting, the uncertainty can scale like $1/N$, which saturates the corresponding Cream\'er-Rao bound across the whole parameter regime $\varphi \in [-\pi, \pi)$. In the presence of photon loss, the Fisher information can still follow the Heisenberg scaling, while the uncertainty of the phase estimation is no longer parameter-independent but can saturate the Cream\'er-Rao bound for some parameters. We have shown that we can recover the parameter-independent uncertainty even in the presence of photon loss by means of a projective measurement on the ancilla two-level system. The outcome of this probe will determine if the protocol may proceed or a potential restart of the process is needed. The likelihood of restarting the measurement process can be lowered if the photon loss is small. The  parameter estimation protocol presented here may be extendable to other types of parameters or multiple parameters. It is also of interest to consider other open systems such as phase damping channels and classical noises.

\section{Acknowledgement}

This work is supported by the ART020-Quantum Technologies Project. 




\end{document}